\documentclass[twoside]{ilcws07}
\usepackage[latin1]{inputenc}
\usepackage[dvips]{graphicx,epsfig,color}
\usepackage{wrapfig,rotating}
\usepackage{amssymb,amsmath,array}

\begin{document}
\title{
Particle Tracking with a Thin Pixel Telescope} 
\author{Marco Battaglia$^{1,2}$,
Devis Contarato$^2$, 
Piero Giubilato$^{2,3}$,\\
Lindsay Glesener$^2$,
Leo Greiner$^2$,
Benjamin Hooberman$^2$
\vspace{.3cm}\\
1- University of California - Dept of Physics, Berkeley, CA - USA \\
\vspace*{-.2cm}\\
2- Lawrence Berkeley National Laboratory, Berkeley, CA - USA \\
\vspace*{-.2cm}\\
3- Istituto Nazionale di Fisica Nucleare and Dip. di Fisica, Padova, Italy \\
}

\maketitle

\begin{abstract}
We report results on a tracking performance study performed using a beam 
telescope made of 50~$\mu$m-thick CMOS pixel sensors on the 1.5~GeV electron 
beam at the LBNL ALS.

\end{abstract}

\section{The Thin Pixel Prototype Telescope}

The anticipated ILC physics program indicates that identification of 
heavy fermions with high efficiency and purity is of primary importance.
The requirements in terms of track extrapolation accuracy are 
$\simeq 5~\mu{\mathrm{m}} \oplus \frac{10~\mu{\mathrm{m}}}{{\mathrm{p_t~(GeV)}}}$.
The Vertex Tracker design and the sensor R\&D is driven by this requirement.
CMOS pixel sensors are an attractive option for the ILC Vertex Tracker. 
In particular, they can be back-thinned to 50~$\mu$m or less, without  
significant performance deterioration~\cite{Battaglia:2006tf}.

While there has already been some experience with the reconstruction of 
well-isolated, high momentum particles with monolithic pixel 
sensors~\cite{Stanic:2006da,Trimpl:2006dc}, no data exists on low momentum
 particle tracking with thin pixel sensors with occupancy conditions 
comparable to those expected at the ILC. We present here results obtained 
with a Thin Pixel Prototype Telescope (TPPT) on the 1.5~GeV 
$e^-$ beam at the LBNL Advanced Light Source (ALS) accelerator complex.      
The TPPT is the first beam telescope made of thinned CMOS pixel sensors. 
It consists of three planes of thin pixel sensors (layers 1 to 3), each spaced 
by 17~mm. One additional detector (layer 4) is added 17~mm downstream of 
the third layer. The MIMOSA~5 chip~\cite{mimosa, deptuch-ref}, developed at IPHC in 
Strasbourg, France, has been selected for the TPPT. This chip, fabricated in the 
0.6~$\mu$m AMS process, features a large active area of 1.7$\times$1.7~cm$^2$ and 
more than 1~M pixels. The epitaxial layer is 14~$\mu$m thick and the pixel pitch 
17~$\mu$m.One sector of each MIMOSA~5 chip, corresponding to a 510$\times$512 pixel 
array, is readout through a custom FPGA-driven acquisition board. 
The beam spill consists of a single bunch with a repetition rate of 1~Hz and the 
extraction signal is used as a trigger. For each spill we acquire three frames, one 
before the particles arrive on the TPPT. Four 14~bits, 40~MSample/s ADCs simultaneously read 
the four sensors, while an array of digital buffers drive all the
required clocks and synchronisation signals. A 32~bits wide bus connects the FPGA 
to a digital acquisition board installed on a control PC. Data is processed on-line 
to perform correlated double sampling, pedestal subtraction, noise computation and 
cluster identification.  To reduce the amount of data written to disk only the 
addresses and pulse heights of the pixels in a fixed matrix around the centre of a 
cluster candidate are recorded. The data is then converted in the {\tt lcio} format 
and the offline analysis is performed by processors developed within 
the {\tt Marlin} framework~\cite{Gaede:2006pj}. Each event is scanned for pixels 
with pulse height over a signal-to-noise (S/N) threshold of 4.5, these are designated 
as cluster `seeds'. Seeds are then sorted according to their pulse height values and 
the surrounding, neighbouring pixels are tested for addition to the cluster. Pixels 
with a pulse height in excess to 2.0 time the noise are accepted. The neighbour 
search is performed in a 5$\times$5 matrix. Clusters are not allowed to overlap and 
we require that clusters are not discontinuous, i.e. pixels associated to a cluster 
cannot be interleaved by any pixel below the neighbour threshold. The point of 
impact of the particle track on the detector is determined by reconstructing the charge
centre-of-gravity of the cluster.

\section{The ALS Beam Test with 1.5~GeV Electrons}
 
The TPPT telescope is operated in an optical enclosure mounted on an optical rail and 
aligned on the ALS BTF beam line. We report here results of the first data taking 
performed in Fall~2006. The temperature was kept constant during operation 
at $\simeq$~27$^{\circ}$C by forced airflow. The cluster S/N averaged on the TPPT layers 
is 14.5. Due to a drop of the voltage supplied to the chip, one of the planes of the 
telescope was not fully efficient during the ALS data taking. Because of that we also 
accept particle tracks reconstructed on two layers, provided they have a hit on the 
reference plane within 150~$\mu$m.
\begin{wrapfigure}{c}{0.5\columnwidth}
\centerline{\includegraphics[width=0.46\columnwidth]{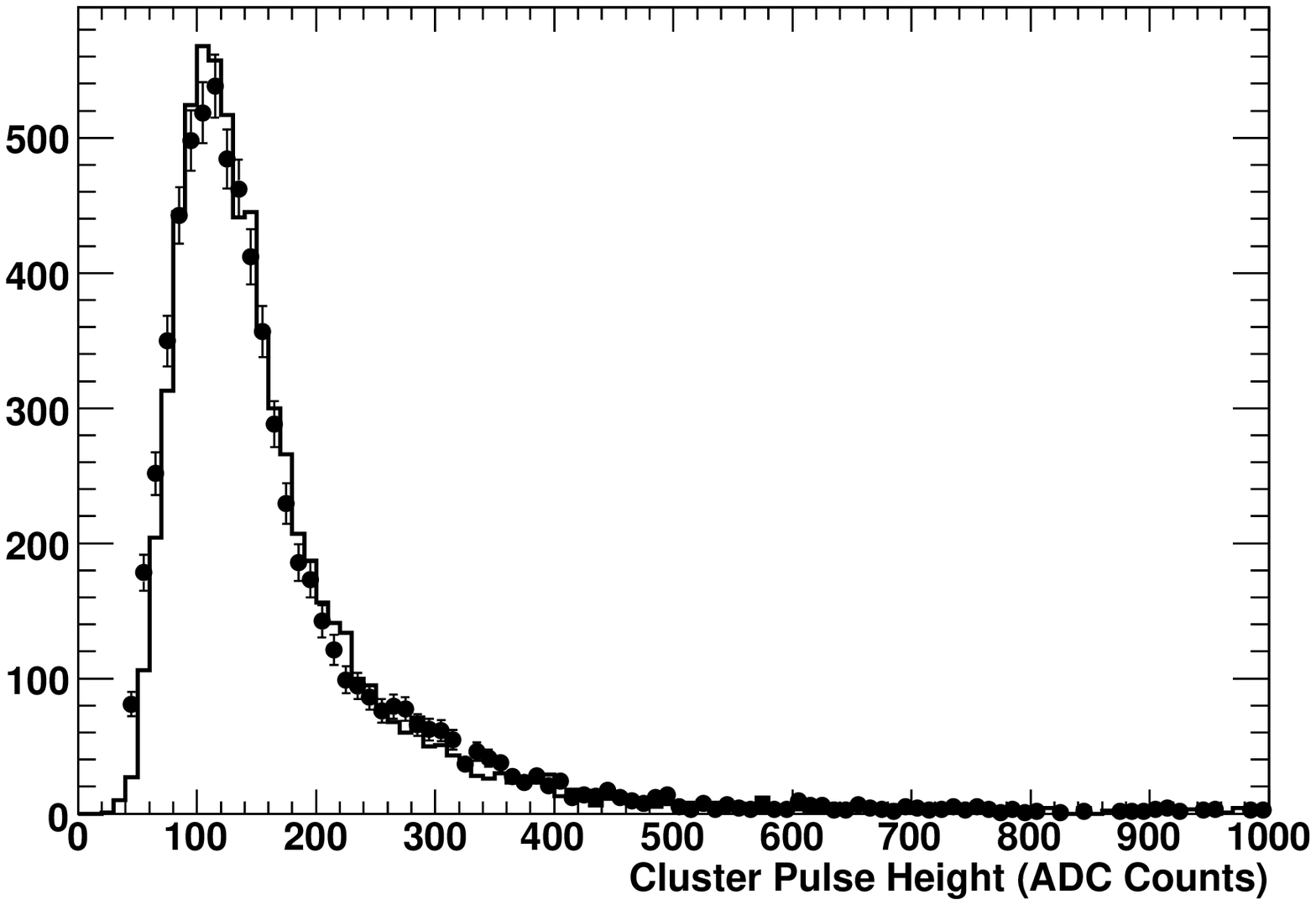}}
\centerline{\includegraphics[width=0.46\columnwidth]{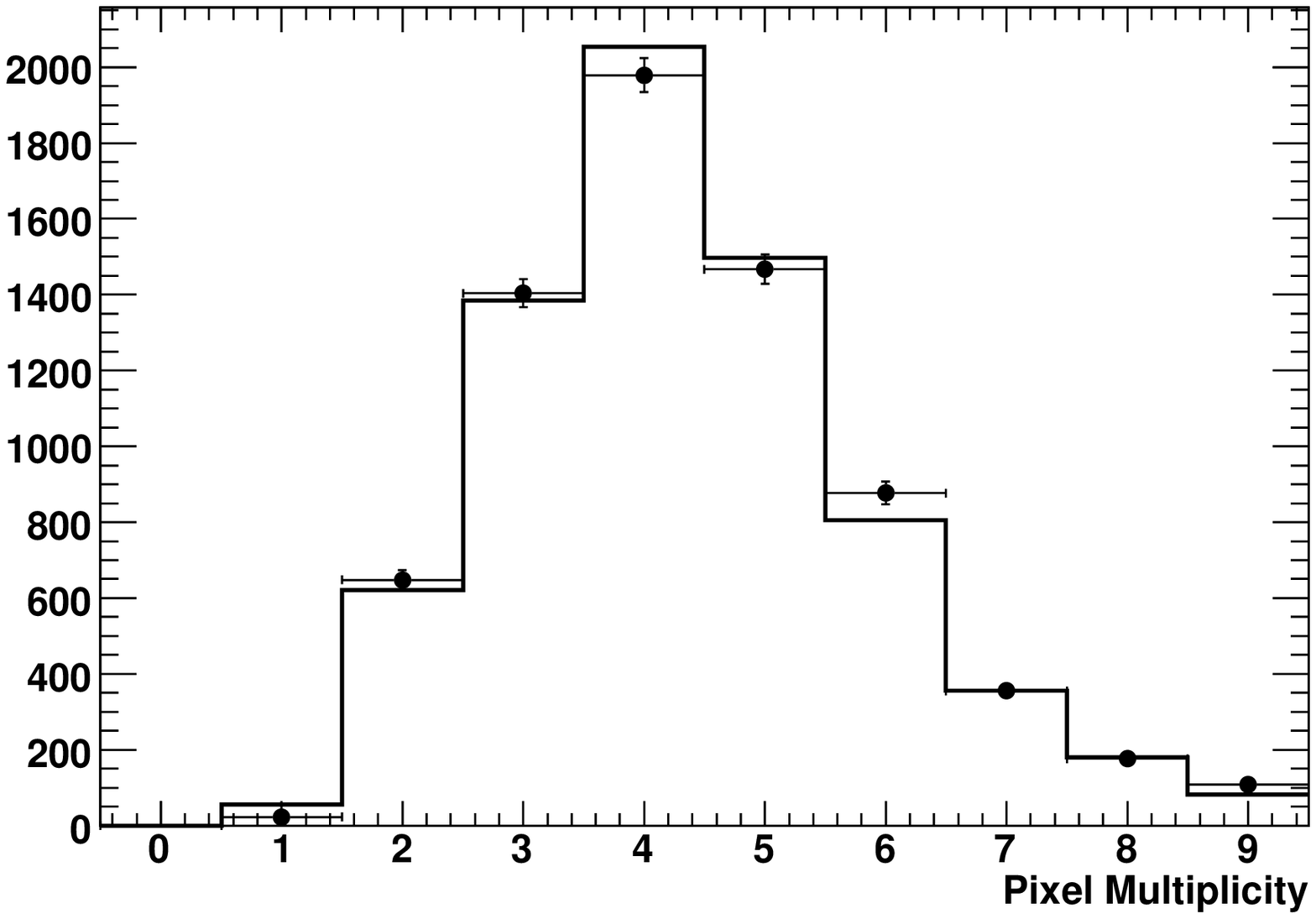}}
\caption{Detector response to 1.5~GeV $e^-$s in data (point with error bars) and 
in the {\tt Geant-4}+{\tt Marlin} simulation and reconstruction. The cluster signal 
pulse height (left) and pixel multiplicity in a cluster (right) are shown for clusters 
associated to reconstructed particle tracks.}
\label{Fig:1}
\end{wrapfigure}
Detailed simulation of the TPPT has been carried out, using the
{\tt Geant-4} package~\cite{Agostinelli:2002hh} to generate the 
particle points of impact and energy deposits on the sensitive planes.
These have been stored in {\tt lcio} format and used as input to 
a CMOS pixel simulation program implemented in {\tt Marlin}~\cite{Battaglia:2007eu}. 
Noise values have been matched to those measured for the detectors in 
the telescope. Digitised simulation has been passed through the same 
cluster reconstruction  program as the real data and reconstructed hits used 
to fit straight particle tracks, after pattern recognition. The single detector 
response is well reproduced by the simulation. Figure~\ref{Fig:1} compares the 
cluster pulse height and the pixels multiplicity in a cluster for simulation and 
ALS beam test data.

The TPPT at the ALS beam-line allows us to perform detailed studies of particle tracking 
with various, controllable, levels of track density under realistic conditions.
Data have been collected at the BTS with different beam intensities ranging 
from 0.5 particles~mm$^{-2}$ up to about 5 particles~mm$^{-2}$. These particle 
densities resemble those expected in the core of hadronic jets at $\sqrt{s}$ = 500~GeV.
In particular, the distribution of the distance between a hit associated to a particle 
track and its closest hit reconstructed on the same layer in the high intensity runs at 
the ALS reproduces that predicted by simulation for 
$e^+e^- \to Z^-H^0 \to q \bar q b \bar b$ for $M_H$ = 120~GeV at 500~GeV.

After data taking, the beam telescope geometry has been surveyed using an optical
metrology machine. The results of this survey have been used as starting point
of the alignment procedure, performed on a sample of approximately 20000 
well-isolated particle tracks with four correlated hits. After alignment, 
track candidates have been defined by matching hits on two layers. 
Each track candidate has been extrapolated to the third layer, where 
the closest hit has been added, if its residual was less than 50~$\mu$m.
Pattern recognition ambiguities have been solved based on the number of 
associated hits and on the difference between the track slope 
and the expected beam slope, determined from the settings of the beam-line final dipoles. 
Tracks have then been re-fitted using a modified least-square, to account for kinks due 
to multiple scattering on the measuring planes~\cite{Lutz:1987vs} and extrapolated to 
the reference layer, where the residual to the closest reconstructed hit has been computed. 
Since the beam contains a fraction of lower momentum particles, which increases 
moving away from the beam axis, only the central region of the beam was used and an 
additional cut was imposed on the residual on the second coordinate. 
\begin{wrapfigure}{r}{0.70\columnwidth}
\centerline{\includegraphics[width=0.69\columnwidth]{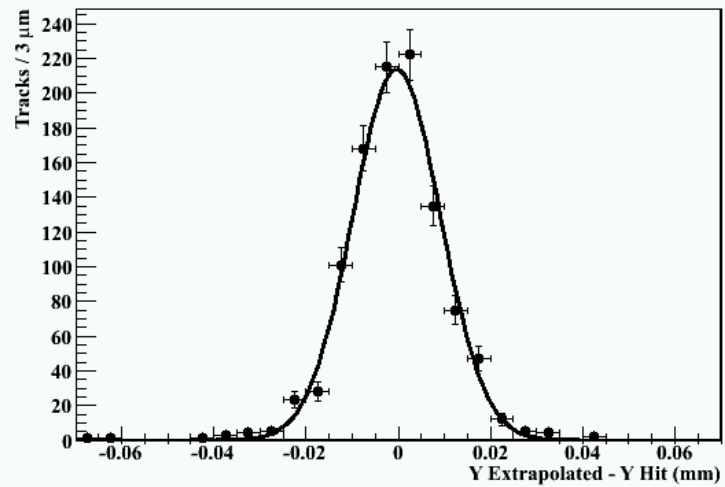}}
\caption{Residual on the vertical coordinate between the extrapolated track position 
on TPPT first layer and the associated hit reconstructed coordinate. The fitted 
Gaussian curve has a width of 9.4~$\mu$m.}
\label{Fig:2}
\end{wrapfigure} 
The tracking performance 
has been characterised by the residuals between the extrapolation of a track reconstructed 
using at least two layers and the point reconstructed on the reference layer using two 
different geometries. 
The first uses the extrapolation of the track back to the first layer. This 
resembles closely the case of track extrapolation from the vertex detector to the interaction 
point. Using all accepted tracks we measure a residual of (9.4$\pm$0.2)~$\mu$m (see Figure~\ref{Fig:2}), 
which becomes (8.9$\pm$0.4)~$\mu$m when restricting to three-hit tracks. This should be compared to 
6.8~$\mu$m obtained from the 
simulation, which assumes perfect geometry. Subtracting in quadrature the estimated single point 
resolution of x~$\mu$m, we obtain an extrapolation resolution of 8.5~$\mu$m for a 1.5~GeV 
particle, which is consistent with the impact parameter resolution required for the
ILC. The second geometry adopted extrapolates the track on the second detector layer. 
In this case the multiple scattering effect is reduced by having measurements on both 
side of the extrapolation plane and we measure a residual of (6.9$\pm$0.1)~$\mu$m.

Despite the multiple scattering effect, the extrapolation resolution of the TPPT at 
the ALS is significantly smaller than the MIMOSA-5 pixel pitch. This allows to perform 
studies of cluster shape as a function of the track point of impact. We compared the 
number of pixels, along the horizontal coordinate, in clusters reconstructed on the 
first layer which are associated to two sets of tracks. In the first set tracks 
with extrapolation along the $x$ horizontal axis within 4~$\mu$m from the pixel centre are 
chosen: the average pixel multiplicity is 1.5. In the second set the track intercepts the 
detector more than 8~$\mu$m away from the pixel centre: the average pixel multiplicity 
increases to 2.3. 

\section{The T-966 Telescope}

A second thin pixel telescope (TPPT-2) has been built for use in the T-966 beam test 
experiment of the 120~GeV proton beam at the Fermilab MTest facility. The TPPT-2 
consists of four layers of 50~$\mu$m-thin MIMOSA-5 sensors mounted on new mezzanine 
cards with low profile components and larger clear region in the PC board below chip.
The four layers are mounted using precision mechanics and are spaced by 15~mm. 
The chips have been positioned on the mezzanine boards using a precision vacuum chuck
which gives a mounting accuracy better than 50~$\mu$m. Downstream from the TPPT-2, 
a detector under test (DUT) can be mounted on a computer-controlled XY stage which 
allows to remotely align it to the telescope. The DUT spacing from the TPPT-2 can be 
varied from 5~mm to 20~mm. The TPPT-2 has been tested in May~2007 on a 1.23~GeV $e^-$ 
beam extracted from the ALS. An average S/N of 15.5 has been measured operating at 
27$^{\circ}$C. For the first T-966 data taking, planned for July and August 2007, an 
operating temperature of 20$^{\circ}$C is foreseen, which will reduce the noise. 




\section*{Acknowledgements}

This work was supported by the Director, Office of Science, of the 
U.S. Department of Energy under Contract 
No.DE-AC02-05CH11231 and used resources of the National Energy Research Scientific 
Computing Center, supported under Contract No.DE-AC03-76SF00098.
We are indebted to the staff of the LBNL Advanced Light Source for their 
help and the excellent performance of the machine.


\begin{footnotesize}

\end{footnotesize}
\end{document}